\documentstyle[sprocl,epsf]{article}

\def\gsim{\ \rlap{\raise 3pt \hbox{$>$}}{\lower 3pt \hbox{$\sim$}}\ }
\def\lsim{\ \rlap{\raise 3pt \hbox{$<$}}{\lower 3pt \hbox{$\sim$}}\ }

\begin{document}

%%% start Cornell preprint title page %%%%%%%%%%%%%
\begin{titlepage}

\begin{flushright}
CLNS~00/1685\\
August 2000\\[0.15cm]
{\tt hep-ph/0008072}
\end{flushright}

\vspace{1.5cm}

\begin{center}
\Large\bf 
Applications of QCD Factorization in\\ 
Hadronic \boldmath$B$\unboldmath\ Decays
\end{center}

\vspace{1.2cm}

\begin{center}
Matthias Neubert\\
{\sl Newman Laboratory of Nuclear Studies, Cornell University\\
Ithaca, New York 14853, U.S.A.}
\end{center}

\vspace{1.3cm}

\begin{center}
{\bf Abstract:}\\[0.3cm]
\parbox{11cm}{
We review recent advances in the theory of strong-interaction effects
and final-state interactions in hadronic weak decays of heavy mesons. 
In the heavy-quark limit, the amplitudes for most nonleptonic, 
two-body $B$ decays can be calculated from first principles and 
expressed in terms of semileptonic form factors and light-cone 
distribution amplitudes. We summarize the main features of this novel 
QCD factorization and discuss its phenomenological applications to
$B\to D^{(*)} L$ decays (with $L$ a light meson), and to the rare 
charmless decays $B\to\pi K$ and $B\to\pi\pi$.}
\end{center}

\vspace{1cm}

\begin{center}
{\sl To appear in the Proceedings of the\\
Fourth Workshop on Continuous Advances in QCD\\
Minneapolis, Minnesota, 12--14 May 2000}
\end{center}
\vfil

\end{titlepage}

\thispagestyle{empty}
\vbox{}
\newpage

\setcounter{page}{1}

%%% end Cornell preprint title page %%%%%%%%%%%%%

\title{APPLICATIONS OF QCD FACTORIZATION IN\\ 
HADRONIC \boldmath$B$\unboldmath\ DECAYS}

\author{M. NEUBERT}

\address{Newman Laboratory of Nuclear Studies, Cornell University\\
Ithaca, New York 14853, U.S.A.}

\maketitle\abstracts{
We review recent advances in the theory of strong-interaction effects
and final-state interactions in hadronic weak decays of heavy mesons. 
In the heavy-quark limit, the amplitudes for most nonleptonic, 
two-body $B$ decays can be calculated from first principles and 
expressed in terms of semileptonic form factors and light-cone 
distribution amplitudes. We summarize the main features of this novel 
QCD factorization and discuss its phenomenological applications to
$B\to D^{(*)} L$ decays (with $L$ a light meson), and to the rare 
charmless decays $B\to\pi K$ and $B\to\pi\pi$.}

\section{Introduction}

\begin{figure}
\epsfxsize=5.0cm
\epsffile{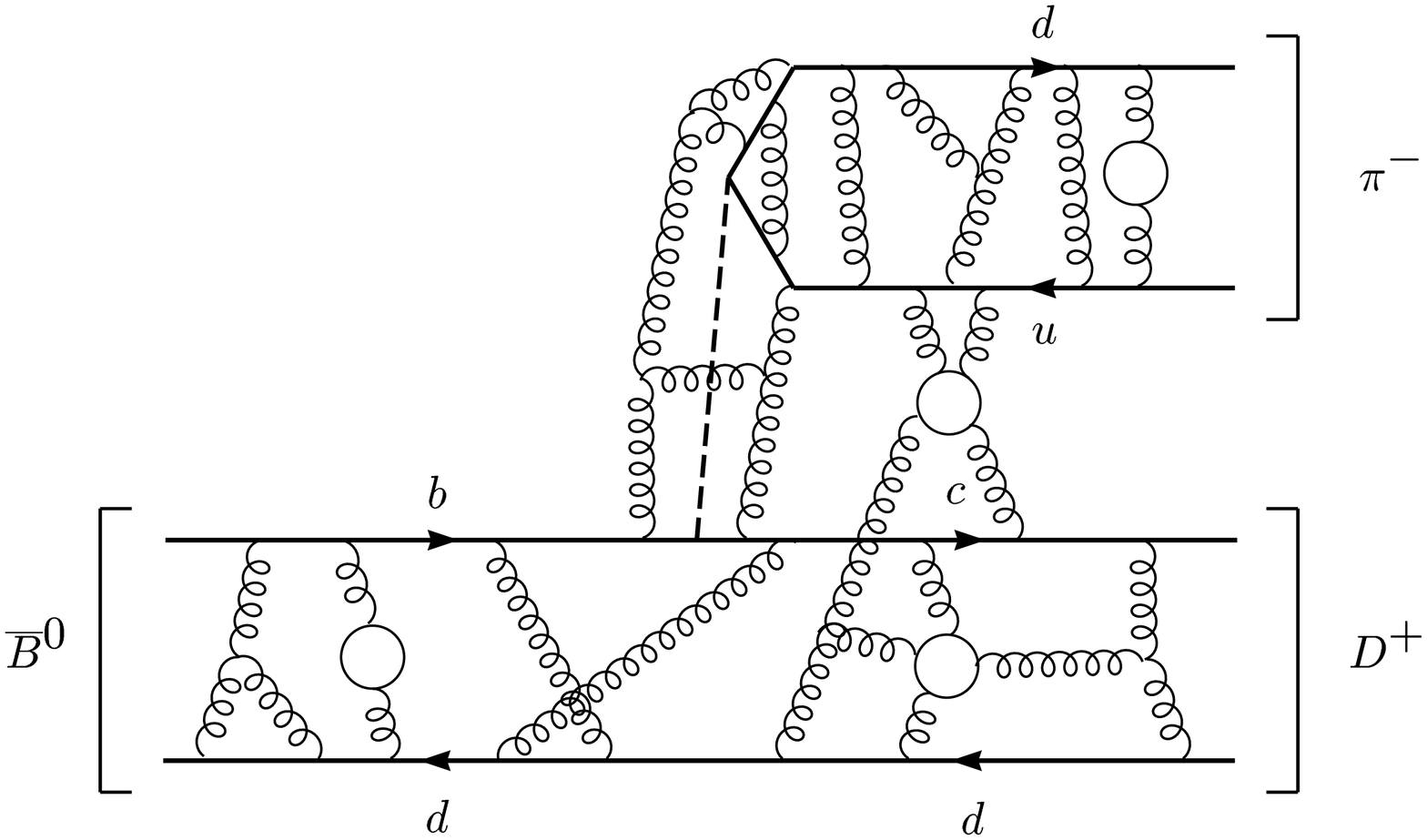}
\epsfxsize=6.7cm
\epsffile{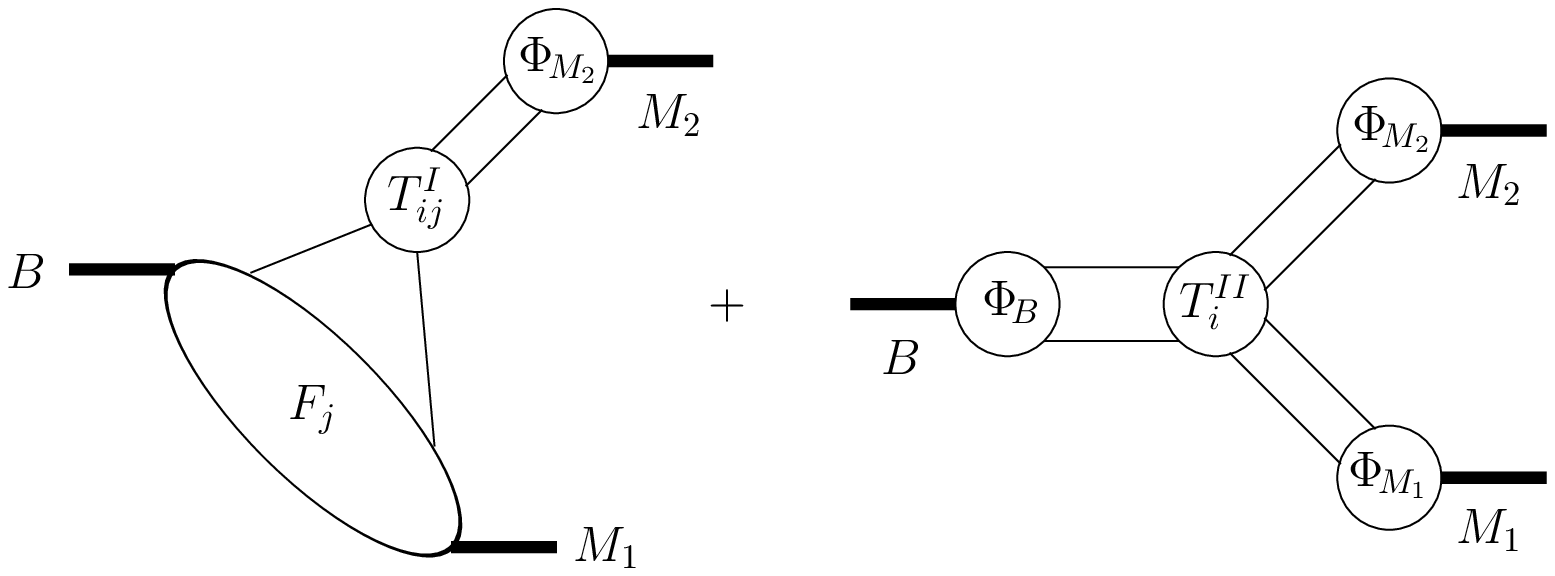}
\caption{Left: Strong-interaction effects in a hadronic weak decay.
Right: QCD factorization in the heavy-quark limit. The second term
is power suppressed for $B\to D\pi$, but must be kept for decays 
with two light mesons in the final state, such as $B\to\pi K$.
Contributions not shown (such as weak annihilation graphs) are power
suppressed.  
\label{fig:nonlep}}
\end{figure}

The theoretical description of hadronic weak decays is difficult due 
to nonperturbative strong-interaction dynamics. This affects the 
interpretation of data collected at the $B$ factories and in many 
cases limits our ability to uncover the origin of CP violation and 
search for New Physics. The complexity of the 
problem is illustrated in the cartoon on the left-hand side of 
Fig.~\ref{fig:nonlep}. 

It is well known how to control the effects of hard gluons with 
virtuality between the electroweak scale $M_W$ and the scale $m_B$
characteristic to the decays of interest. They can be dealt with by 
constructing a low-energy effective weak Hamiltonian 
\begin{equation}\label{Heff}
   {\cal H}_{\rm eff} = \frac{G_F}{\sqrt2} \sum_i
   \lambda_i^{\rm CKM}\,C_i(M_W/\mu)\,O_i(\mu) + \mbox{h.c.},
\end{equation}
where $\lambda_i^{\rm CKM}$ are products of CKM matrix elements, 
$C_i(M_W/\mu)$ are calculable short-distance coefficients, and 
$O_i(\mu)$ are local operators renormalized at a scale 
$\mu={\cal O}(m_B)$. The challenge is to calculate the hadronic 
matrix elements of these operators with controlled theoretical 
uncertainties, using a systematic approximation scheme.

Previous field-theoretic attempts to evaluate these matrix elements 
have employed dynamical schemes such as lattice field theory, QCD sum 
rules, or the hard-scattering approach. The first two have
difficulties in accounting for final-state rescattering, which however 
is important for predicting direct CP asymmetries. The hard-scattering 
approach misses the leading soft contribution to the 
$B\to\mbox{meson}$ transition form factors and thus falls short of 
reproducing the correct magnitude of the decay amplitudes. In view 
of these difficulties, most previous analyses of hadronic decays have 
employed phenomenological models such as ``naive'' or ``generalized 
factorization'', in which the complicated matrix elements of four-quark 
operators in the effective weak Hamiltonian are replaced, in an {\em 
ad hoc\/} way, by products of current matrix elements. Corrections to 
this approximation are accounted for by introducing a set of 
phenomenological parameters $a_i$. A different strategy is to classify 
nonleptonic decay amplitudes according to flavor topologies (``trees'' 
and ``penguins''), which can be decomposed into SU(3) or isospin 
amplitudes. This leads to relations between decay amplitudes in the 
flavor-symmetry limit. No attempt is made, however, to compute these 
amplitudes from first principles.

\section{QCD Factorization Formula}

Here we summarize recent progress in the theoretical understanding
of nonleptonic decay amplitudes in the heavy-quark 
limit.\cite{BBNS,bigpaper} The underlying idea is to exploit the 
presence of a large scale, i.e., the fact that 
$m_b\gg\Lambda_{\rm QCD}$. In order to disentangle the physics 
associated with these two scales, we factorize and compute hard 
contributions to the decay amplitudes arising from gluons with 
virtuality of order $m_b$, and parameterize soft and collinear 
contributions. Considering the cartoon in Fig.~\ref{fig:nonlep}, we 
denote by $M_1$ the meson that absorbs the spectator quark of the $B$ 
meson, and by $M_2$ the meson at the upper vertex, to which we refer 
as the ``emission particle''. We find that at leading power in 
$\Lambda_{\rm QCD}/m_b$ all long-distance contributions to the decay 
amplitudes can be factorized into semileptonic form factors and meson 
light-cone distribution amplitudes, which are much simpler quantities 
than the nonleptonic amplitudes themselves. A graphical representation 
of the resulting ``factorization formula'' is shown on the right-hand 
side in Fig.~\ref{fig:nonlep}. The physical picture underlying 
factorization is color transparency.\cite{Bj89,DG91} If the emission 
particle is a light meson, its constituents carry large energy of 
order $m_b$ and are nearly collinear. Soft gluons coupling to this 
system see only its net zero color charge and hence decouple. 
Interactions with the color dipole of the small $q\bar q$-pair are 
power suppressed in the heavy-quark limit.

For $B$ decays into final states containing a heavy charm meson and 
a light meson the factorization formula takes the form
\begin{equation}\label{fff}
   \langle D^{(*)+} L^-|\,O_i\,|\bar B_d\rangle
   = \sum_j F_j^{B\to D^{(*)}} f_L\!
   \int\limits_0^1\!\mbox{d}u\,T_{ij}^{\rm I}(u)\,\Phi_L(u)
   + {\cal O}\bigg(\frac{\Lambda_{\rm QCD}}{m_b}\bigg) \,,
\end{equation}
where $O_i$ is an operator in the effective weak Hamiltonian 
(\ref{Heff}), $F_j^{B\to D^{(*)}}$ are transition form factors 
(evaluated at $q^2=m_L^2\approx 0$), $f_L$ and $\Phi_L$ are the decay 
constant and leading-twist light-cone distribution amplitude of the 
light meson, and $T_{ij}^{\rm I}$ are process-dependent 
hard-scattering kernels. For 
decays into final states containing two light mesons there is a second 
type of contribution to the factorization formula, which involves a 
hard interaction with the spectator quark in the $B$ meson. This is 
shown by the second graph on the right-hand side in 
Fig.~\ref{fig:nonlep}. Below we focus first on $\bar B\to D^{(*)} L$ 
decays (with $L$ a light meson), where this second term is power 
suppressed 
and can be neglected. Decays into two light final-state mesons are 
more complicated~\cite{BBNS,BBNSnew} and will be discussed briefly in 
Sect.~\ref{sec:piK}. A more detailed account of the conceptual 
foundations of the QCD factorization approach has been presented by 
Beneke at this Workshop~\cite{Martin}.

The factorization formula for nonleptonic decays provides a 
model-inde\-pen\-dent basis for the analysis of these processes in an 
expansion in powers and logarithms of $\Lambda_{\rm QCD}/m_b$. At
leading power, but to all orders in $\alpha_s$, the decay amplitudes
assume the factorized form shown in (\ref{fff}). Having such a
formalism based on power counting in $\Lambda_{\rm QCD}/m_b$ is of 
great importance to the theoretical description of hadronic weak 
decays, since it provides a well-defined limit of QCD in which these 
processes admit a rigorous, theoretical description. (For instance, 
the possibility to compute systematically $O(\alpha_s)$ corrections to 
``naive factorization'', which emerges as the leading term in the 
heavy-quark limit, solves the old problem of renormalization-scale and 
scheme dependences of nonleptonic amplitudes.) The usefulness of
this new scheme may be compared with the usefulness of the heavy-quark
effective theory for the analysis of exclusive semileptonic decays of
heavy mesons, or of the heavy-quark expansion for the analysis of
inclusive decay rates. In all three cases, it is the fact that 
hadronic uncertainties can be eliminated up to power corrections
in $\Lambda_{\rm QCD}/m_b$ that has advanced our ability to control 
theoretical errors.

It must be stressed, however, that we are just beginning to explore
the theory of nonleptonic $B$ decays. Some important conceptual 
problems remain to be better understood. In the next few years it 
will be important to further develop the approach. This should 
include an all-orders proof of factorization at leading power, the 
development of a formalism for dealing with power corrections to 
factorization, understanding the light-cone structure of heavy mesons, 
and understanding the relevance (or irrelevance) of Sudakov form 
factors. Also, we must gauge the accuracy of the approach by
learning about the magnitude of corrections to the heavy-quark limit
from extensive comparisons of theoretical predictions with data. 

As experience with previous heavy-quark expansions has shown, this 
is going to be a long route. Yet, already we have obtained important 
insights. Before turning to specific applications, let us mention 
three points here:

1. Corrections to ``naive factorization'' (usually called 
``nonfactorizable effects'') are process dependent, in contrast with 
a basic assumption underlying models of ``generalized factorization''. 

2. The physics of nonleptonic decays is both rich and complicated. 
There may, in general, be an interplay of several small parameters 
(Wilson coefficients, CKM factors, $1/N_c$, etc.) in addition to the 
small parameter $\Lambda_{\rm QCD}/m_b$ relevant to QCD factorization. 
Also, several not-so-well-known input parameters (e.g., heavy-to-light 
form factors and light-cone distribution amplitudes) introducing 
numerical uncertainties in the predictions.

3. Strong-interaction phases arising from final-state interactions are
suppressed in the heavy-quark limit. More precisely, the imaginary 
parts of nonleptonic decay amplitudes are suppressed by at least one 
power of $\alpha_s(m_b)$ or $\Lambda_{\rm QCD}/m_b$. At leading power, 
the phases are calculable from the imaginary parts of the 
hard-scattering kernels in the factorization formula.

\section{\boldmath
Applications to $\bar B_d\to D^{(*)+} L^-$ Decays
\unboldmath}

Our result for the nonleptonic $\bar B_d\to D^{(*)+}L^-$ decay 
amplitudes (with $L$ a light meson) can be compactly expressed in 
terms of the matrix elements of a ``transition operator''
\begin{equation}\label{heffa1}
   {\cal T} = \frac{G_F}{\sqrt2}\,V^*_{ud} V_{cb}
   \Big[ a_1(D L)\,Q_V - a_1(D^* L)\,Q_A \Big] \,,
\end{equation}
where the hadronic matrix elements of the operators 
$Q_V=\bar c\gamma^\mu b\,\otimes\,\bar d\gamma_\mu(1-\gamma_5)u$
and $Q_A=\bar c\gamma^\mu\gamma_5 b\,\otimes\,
\bar d\gamma_\mu(1-\gamma_5)u$ are understood to be evaluated in 
factorized form. Eq.~(\ref{heffa1}) defines the quantities 
$a_1(D^{(*)} L)$, which include the leading ``nonfactorizable'' 
corrections, in a renormalization-group invariant way. To leading power 
in $\Lambda_{\rm QCD}/m_b$ these quantities should not be interpreted as 
phenomenological parameters (as is usually done), because they are 
dominated by hard gluon exchange and thus calculable in QCD. At 
next-to-leading order in $\alpha_s$ we obtain~\cite{bigpaper}
\begin{equation}\label{a1}
   a_1(D^{(*)}L) = \bar C_1(m_b) + \frac{\bar C_2(m_b)}{N_c} \left[
   1 + \frac{C_F\alpha_s(m_b)}{4\pi} \int\limits^1_0\!\mbox{d}u\,
   F(u,\pm z)\,\Phi_L(u) \right] \,,
\end{equation}
where $\bar C_i(m_b)$ are the so-called ``renormalization-scheme 
independent'' Wilson coefficients, and the upper (lower) sign in the 
second argument of the function $F(u,\pm z)$ refers to a $D$ ($D^*$) 
meson in the final state. The exact expression for this 
function is known but not relevant to our discussion here. Note that 
the coefficients $a_1(D L)$ and $a_1(D^* L)$ are nonuniversal, i.e., 
they are explicitly dependent on the nature of the final-state mesons. 
Politzer and Wise have computed the ``nonfactorizable'' vertex 
corrections to the decay rate ratio of $\bar B_d\to D^+\pi^-$ and 
$\bar B_d\to D^{*+}\pi^-$.\cite{PW91} This requires the symmetric part 
(with respect to $u\leftrightarrow 1-u$) of the difference 
$F(u,z)-F(u,-z)$. We agree with their result.

The expressions for the decay amplitudes obtained by evaluating the 
had\-ro\-nic matrix elements of the transition operator ${\cal T}$ 
involve products of CKM matrix elements, light-meson decay constants, 
$\bar B\to D^{(*)}$ transition form factors, and the QCD parameters 
$a_1(D^{(*)} L)$. A numerical analysis shows that 
$|a_1|=1.055\pm 0.025$ for the decays considered below. Below we will 
use this as our central value.

\subsection{Tests of factorization}

A particularly clean test of our predictions is obtained by relating 
the $\bar B_d\to D^{*+} L^-$ decay rates to the differential 
semileptonic $\bar B_d\to D^{*+}\,l^-\nu$ decay rate evaluated at 
$q^2=m_L^2$. In this way the parameters $|a_1|$ can be measured 
directly, since~\cite{Bj89}
\begin{equation}
   \frac{\Gamma(\bar B_d\to D^{*+} L^-)}
    {d\Gamma(\bar B_d\to D^{*+} l^-\bar\nu)/dq^2\big|_{q^2=m^2_L}}
   = 6\pi^2 |V_{ud}|^2 f^2_L\,|a_1(D^* L)|^2 \,.
\end{equation}
With our result for $a_1$ this relation becomes a prediction based on 
first principles of QCD. This is to be contrasted with the usual 
interpretation of this formula, where $a_1$ plays the role of a 
phenomenological parameter that is fitted from data.

Using results reported by the CLEO Collaboration,\cite{Rodr97} we find
\begin{equation}
   |a_1(D^*\pi)| = 1.08 \pm 0.07 \,, \quad
   |a_1(D^*\rho)| = 1.09\pm 0.10 \,, \quad
   |a_1(D^* a_1)| = 1.08\pm 0.11 \,,
\end{equation}
in good agreement with our prediction. It is reassuring that the data 
show no evidence for large power corrections to our results. However, 
a further improvement in the experimental accuracy would be desirable 
in order to become sensitive to process-dependent, nonfactorizable 
effects.

\subsection{Predictions for class-I decay amplitudes}

We now consider a larger set of so-called class-I decays of the form 
$\bar B_d\to D^{(*)+} L^-$, all of which are governed by the transition 
operator (\ref{heffa1}). In Tab.~\ref{tab:10decays} we compare
the QCD factorization predictions with experimental data. As previously 
we work in the heavy-quark limit, i.e., our predictions are model 
independent up to corrections suppressed by at least one power of 
$\Lambda_{\rm QCD}/m_b$. There is good agreement between our predictions 
and the data within experimental errors, which however are still large. 
It would be desirable to reduce these errors to the percent level. 
(Note that we have not attempted to adjust the semileptonic form 
factors $F_+^{\bar B\to D}$ and $A_0^{\bar B\to D^*}$ entering our 
results so as to obtain a best fit to the data.)

\begin{table}
\caption{\label{tab:10decays}
Model-independent predictions for the branching ratios (in units of
$10^{-3}$) of $\bar B_d\to D^{(*)+} L^-$ decays in the heavy-quark 
limit. Predictions are in units of $(|V_{cb}|/0.04)^2\times
(|a_1|/1.05)^2\times(\tau_{B_d}/1.56\,\mbox{ps})$. We show experimental 
results reported by the CLEO Collaboration~\protect\cite{CLEO9701} and 
the Particle Data Group.\protect\cite{PDG}}
\vspace{0.2cm}
\begin{center}
\begin{tabular}{|l|c|cc|}
\hline
&&&\\[-0.4cm]
Decay mode & Theory (HQL) & CLEO data & PDG98~ \\
&&&\\[-0.4cm]
\hline
&&&\\[-0.4cm]
$\bar B_d\to D^+\pi^-$   & $3.27\times[F_+(0)/0.6]^2$ & $2.50\pm 0.40$
 & $3.0\pm 0.4$ \\
$\bar B_d\to D^+ K^-$    & $0.25\times[F_+(0)/0.6]^2$ & --- & --- \\
$\bar B_d\to D^+\rho^-$  & $7.64\times[F_+(0)/0.6]^2$ & $7.89\pm 1.39$
 & $7.9\pm 1.4$ \\
$\bar B_d\to D^+ K^{*-}$ & $0.39\times[F_+(0)/0.6]^2$ & --- & --- \\
$\bar B_d\to D^+ a_1^-$  & $7.76\times[F_+(0)/0.6]^2$ & $8.34\pm 1.66$
 & $6.0\pm 3.3$ \\
&&&\\[-0.4cm]
\hline
&&&\\[-0.4cm]
$\bar B_d\to D^{*+}\pi^-$   & $3.05\times[A_0(0)/0.6]^2$
 & $2.34\pm 0.32$ & $2.8\pm 0.2$ \\
$\bar B_d\to D^{*+} K^-$    & $0.22\times[A_0(0)/0.6]^2$ & --- & --- \\
$\bar B_d\to D^{*+}\rho^-$  & $7.59\times[A_0(0)/0.6]^2$
 & $7.34\pm 1.00$ & $6.7\pm 3.3$ \\
$\bar B_d\to D^{*+} K^{*-}$ & $0.40\times[A_0(0)/0.6]^2$ & --- & --- \\
$\bar B_d\to D^{*+} a_1^-$  & $8.53\times[A_0(0)/0.6]^2$
 & $11.57\pm 2.02$ & $13.0\pm 2.7$ \\[0.1cm]
\hline 
\end{tabular}
\end{center}
\end{table}

The observation that the experimental data on class-I decays into 
heavy--light final states show good agreement with our predictions may 
be taken as (circumstantial) evidence that in these decays there are no 
unexpectedly large power corrections. In our recent work~\cite{bigpaper} 
we have addressed the important question of power corrections 
theoretically by providing estimates for two sources of power-suppressed 
effects: weak annihilation and spectator interactions. We stress that a 
complete account of power corrections to the heavy-quark limit cannot be 
performed in a systematic way, since these effects are no longer 
dominated by hard gluon exchange. However, we believe that our estimates 
are nevertheless instructive. 

We parameterize the annihilation contribution to the 
$\bar B_d\to D^+\pi^-$ decay amplitude in terms of an amplitude $A$ 
such that ${\cal A}(\bar B_d\to D^+\pi^-)=T+A$, where $T$ is the ``tree 
topology'', which contains the dominant factorizable contribution. We 
find that $A/T\sim 0.04$. We have also obtained an estimate of 
nonfactorizable spectator interactions, which are part of $T$, 
finding that $T_{\rm spec}/T_{\rm lead}\sim-0.03$. In both cases, the 
results exhibit the expected linear power suppression in the 
heavy-quark limit. We conclude that the typical size of power 
corrections in class-I decays into heavy--light final states is at 
the level of 10\% or less, and thus our predictions 
for the values and the near universality of the parameters $a_1$ 
governing these decay modes appear robust.

\section{QCD Factorization in Charmless Hadronic 
\boldmath$B$\unboldmath\ Decays}
\label{sec:piK}

The observation of rare charmless $B$ decays into $\pi K$ and $\pi\pi$ 
final states has resulted in a large amount of theoretical and 
phenomenological work that attempts to interpret these observations in 
terms of the factorization approximation, or in terms of general 
parameterizations of the decay amplitudes. A detailed understanding of 
these amplitudes would help us to pin down the value of the CKM angle 
$\gamma=\mbox{arg}(V_{ub}^*)$ using only data on CP-averaged branching 
fractions. Here we briefly summarize the most important consequences 
of the QCD factorization approach for the $\pi K$ and $\pi \pi$ final 
states.\cite{BBNSnew}

To leading order in an expansion in powers of $\Lambda_{\rm QCD}/m_b$, 
the $B\to\pi K$ matrix elements obey the factorization formula shown
on the right-hand side in Fig.~\ref{fig:nonlep}:
\begin{eqnarray}\label{fact}
   \langle\pi K|\,O_i\,|B\rangle
   &=& F_+^{B\to\pi} f_K\,T_{K,i}^{\rm I}*\Phi_K 
    + F_+^{B\to K} f_\pi\,T_{\pi,i}^{\rm I}*\Phi_\pi \nonumber\\ 
   &&\mbox{}+ f_B f_K f_\pi\,T_i^{\rm II}*\Phi_B*\Phi_K*\Phi_\pi \,,
\end{eqnarray}
where the $*$-products imply an integration over the light-cone 
momentum fractions of the constituent quarks inside the mesons. Our 
results are based on hard-scattering kernels including 
all corrections of order $\alpha_s$. Compared to our previous 
discussion of $B\to\pi\pi$ decays,\cite{BBNS} the present analysis 
incorporates three new ingredients: the matrix elements of 
electroweak penguin operators (for $\pi K$ modes), hard-scattering 
kernels for asymmetric light-cone distributions, and the complete set 
of ``chirally enhanced'' $1/m_b$ corrections.
The second and third items have not been considered in other 
generalizations~\cite{DZ00,Muta} of Ref.~\cite{BBNS} to the $\pi K$ 
final states. The third one, in particular, is essential for 
estimating some of the theoretical uncertainties of the approach. For 
completeness, we note that the predictions from QCD factorization 
differ in essential aspects from those obtained in the conventional 
hard-scattering approach.\cite{KLS00} 

Following Ref.~\cite{BBNS}, we have obtained the coefficients 
$a_i(\pi K)$ (with $i=1,\dots,10$) of the effective, factorized 
``transition operator'' defined analogously to the case of 
$B\to\pi\pi$ decays, but augmented by coefficients related to 
electroweak penguin contributions. Chirally enhanced corrections arise 
from twist-3 two-particle light-cone distribution amplitudes, whose 
normalization involves the quark condensate. The relevant parameter, 
$2\mu_\pi/m_b=-4\langle \bar{q}q\rangle/(f_\pi^2 m_b)$, is formally 
of order $\Lambda_{\rm QCD}/m_b$, but large numerically. The
coefficients $a_6$ and $a_8$ are multiplied by this parameter. There 
are also additional chirally enhanced corrections to the 
spectator-interaction term in (\ref{fact}), which turn out to be the 
more important effect. In both cases, these corrections involve 
logarithmically divergent integrals, which violate factorization. For
instance, for matrix elements of $V-A$ operators the hard spectator 
interaction is proportional to ($\bar u\equiv 1-u$, $\bar v\equiv 1-v$)
\begin{equation}
   \int\limits_0^1\!\frac{du}{\bar u} \frac{dv}{\bar v}\, 
   \Phi_K(u)\left(\Phi_\pi(v)+\frac{2\mu_\pi}{m_b} \frac{\bar u}{u}
   \right) 
\end{equation}
when the spectator quark goes to the pion. (Here we used that the 
twist-3 distribution amplitudes can be taken to be the asymptotic
ones when one neglects twist-3 corrections without the chiral 
enhancement.) The divergence of the $v$-integral in the second term
as $\bar v\to 0$ implies that it is dominated by soft gluon
exchange between the spectator quark and the quarks that form the 
kaon. We therefore treat the divergent integral $X=\int_0^1(dv/\bar v)$
as an unknown parameter (different for the penguin and hard-scattering
contributions), which may in principle be complex owing to soft 
rescattering in higher orders. In our numerical analysis we set 
$X=\ln(m_B/0.35\,\mbox{GeV})+r$, where $r$ is chosen randomly inside a 
circle in the complex plane of radius 3 (``realistic'') or 6 
(``conservative''). Our results also depend on the $B$-meson 
parameter~\cite{BBNS} $\lambda_B$, which we vary between 0.2 and 
0.5\,GeV. Finally, there is in some cases a nonnegligible dependence 
of the coefficients $a_i(\pi K)$ on the renormalization scale, which 
we vary between $m_b/2$ and $2m_b$.

We take $|V_{ub}/V_{cb}|=0.085$ and $m_s(2\,\mbox{GeV})=110\,$MeV as 
fixed input to our analysis, noting that ultimately 
these Standard Model parameters, along with the CP-violating phase 
$\gamma$, might be extracted from a simultaneous fit to the $B\to\pi K$ 
and $B\to\pi\pi$ decay rates. We now summarize our main results.

\subsection{Results on $SU(3)$ breaking}

Bounds on $\gamma$ derived from ratios of CP-averaged $B\to\pi K$ 
decay rates,\cite{FM98,NR98} as well as the determination of $\gamma$ 
using the method of Ref.~\cite{NR}, rely on an estimate of $SU(3)$ 
flavor-symmetry violations. We find that ``nonfactorizable'' 
$SU(3)$-breaking effects (i.e., effects not accounted for by the 
different decay constants and form factors of pions and kaons in the 
conventional factorization approximation) do not exceed the level of a 
few percent.

\subsection{Amplitude parameters}

The approach discussed here allows us to calculate the amplitudes for 
$B\to\pi K$ and $B\to\pi\pi$ decays in terms of form factors 
and light-cone distribution amplitudes. Here we focus on decays
whose branching ratios have already been measured. We write 
\begin{equation}\label{para1}
   {\cal A}(B^0\to\pi^+\pi^-) = T\,\left[
   e^{i\gamma} + (P/T)_{\pi\pi} \right] \,, 
\end{equation}
and parameterize the $B\to\pi K$ amplitudes by~\cite{NR98}
\begin{eqnarray}\label{para}
   {\cal A}(B^+\to\pi^+ K^0) &=& P \left( 1 - \varepsilon_a\,
    e^{i\eta} e^{i\gamma} \right) \,, \nonumber\\
   -\sqrt2\,{\cal A}(B^+\to\pi^0 K^+) &=& P \Big[ 1
    - \varepsilon_a\,e^{i\eta} e^{i\gamma} 
    - \varepsilon_{3/2}\,e^{i\phi} (e^{i\gamma} - q\,e^{i\omega})
    \Big] \,, \nonumber\\
   -{\cal A}(B^0\to\pi^- K^+) &=& P \Big[ 1
    - \varepsilon_a\,e^{i\eta} e^{i\gamma} 
    - \varepsilon_T\,e^{i\phi_T} (e^{i\gamma} - q_C\,e^{i\omega_C})
    \Big] \,, \nonumber\\
   \sqrt2\,{\cal A}(B^0\to\pi^0 K^0) &=& {\cal A}(B^+\to\pi^+ K^0)
    + \sqrt2\,{\cal A}(B^+\to\pi^0 K^+) \nonumber\\
   &&\mbox{}- {\cal A}(B^0\to\pi^- K^+) \,.
\end{eqnarray} 
Table~\ref{tab1} summarizes the numerical values for the amplitude 
parameters for the conservative variation of $X$, and variation of the 
other parameters as explained above. The leading-order results 
correspond to the conventional factorization approximation at the fixed
scale $\mu=m_b$. They are strongly scale dependent. In comparison, the 
scale-dependence of the next-to-leading order results is small, with 
the exception of $q_C\,e^{i \omega_C}$. We stress that the ranges shown 
may overestimate the theoretical uncertainty, since the parameter $X$ 
may ultimately be constrained from a subset of branching fractions. 
This is true, in particular, for the quantity $\varepsilon_{3/2}$ in
Tab.~\ref{tab1}, which can be extracted from data.\cite{NR98}

\begin{table}
\caption{Parameters for the $B\to\pi\pi$ and $B\to\pi K$ decay 
amplitudes as defined in (\protect\ref{para1}) and 
(\protect\ref{para}), for conservative variation of all input 
parameters (see text).}
\label{tab1}
\vspace{0.2cm}
\begin{center}
\begin{tabular}{|c|c|c|} 
\hline 
 \raisebox{0pt}[12pt][6pt]{Parameter} & 
\raisebox{0pt}[12pt][6pt]{Range, NLO} & 
 \raisebox{0pt}[12pt][6pt]{LO}  \\
 \hline
 \raisebox{0pt}[12pt][6pt]{$-\varepsilon_a\,e^{i\eta}$} & 
 \raisebox{0pt}[12pt][6pt]{$(0.017\mbox{--}0.020)\,e^{i\,[13,21]^\circ}$}
 & \raisebox{0pt}[12pt][6pt]{$0.02$}\\
\hline
 \raisebox{0pt}[12pt][6pt]{$\varepsilon_{3/2}\,e^{i\phi}$} & 
 \raisebox{0pt}[12pt][6pt]{$(0.20\mbox{--}0.38)\,e^{i\,[-30,7]^\circ}$}
 & \raisebox{0pt}[12pt][6pt]{$0.36$}\\
\hline
 \raisebox{0pt}[12pt][6pt]{$q\,e^{i\omega}$} & 
 \raisebox{0pt}[12pt][6pt]{$(0.53\mbox{--}0.63)\,e^{i\,[-7,3]^\circ}$}
 & \raisebox{0pt}[12pt][6pt]{$0.64$}\\
\hline
 \raisebox{0pt}[12pt][6pt]{$\varepsilon_T\,e^{i\phi_T}$} & 
 \raisebox{0pt}[12pt][6pt]{$(0.20\mbox{--}0.29)\,e^{i\,[-19,3]^\circ}$}
 & \raisebox{0pt}[12pt][6pt]{$0.33$}\\
\hline
 \raisebox{0pt}[12pt][6pt]{$q_C\,e^{i\omega_C}$} & 
 \raisebox{0pt}[12pt][6pt]{$(0.00\mbox{--}0.22)\,e^{i\,[-180,180]^\circ}$}
 & \raisebox{0pt}[12pt][6pt]{$0.06$}\\
\hline
 \raisebox{0pt}[12pt][6pt]{$(P/T)_{\pi\pi}$} & 
 \raisebox{0pt}[12pt][6pt]{$(0.19\mbox{--}0.29)\,e^{i\,[-1,23]^\circ}$}
 & \raisebox{0pt}[12pt][6pt]{$0.16$}\\
\hline
\end{tabular}
\end{center}
\end{table}

\begin{figure}[t]
\epsfxsize=11.9cm
\epsffile{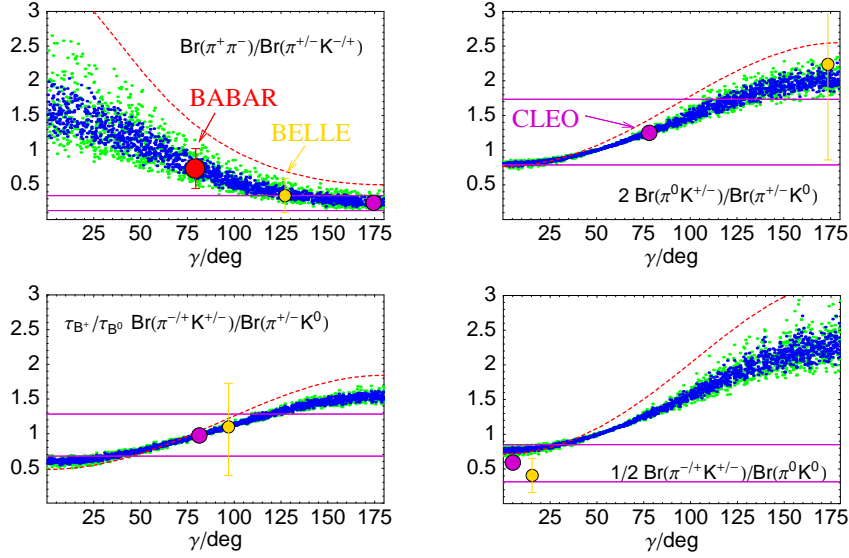}
\caption{Ratios of CP-averaged $B\to\pi K$ and $\pi\pi$ decay rates. 
The scattered points cover a realistic (dark) and conservative (light)
variation of input parameters. The dashed curve is the result obtained
using ``naive factorization''.}
\label{fig1}
\end{figure}

\subsection{Ratios of CP-averaged rates}

Since the relevant form factor $F_+^{B\to\pi}(0)$ is not well known, 
we consider only ratios of CP-averaged branching ratios here. We 
display these as 
functions of the weak phase $\gamma$ in Fig.~\ref{fig1}. For 
comparison, we show the data on the various ratios obtained using
results reported by the CLEO Collaboration~\cite{C00}. We also 
indicate the very recent results reported by the BaBar and Belle 
Collaborations at the ICHEP2000 Conference.\cite{Osaka} From the 
plots and Tab.~\ref{tab1} it is apparent that the corrections with 
respect to 
the conventional factorization approximation are significant (and 
important to reduce the renormalization-scale dependence). Despite 
this fact, the {\em qualitative } pattern that emerges for the set of 
$B\to\pi K$ and $\pi\pi$ decay modes is similar to that in conventional 
factorization. In particular, the penguin--tree interference is 
constructive (destructive) in $B\to\pi^+\pi^-$ ($B\to\pi^- K^+$)
decays if $\gamma<90^\circ$. Taking the currently favored range 
$\gamma=(60\pm 20)^\circ$, we find the following robust predictions:
\begin{eqnarray}\label{rats}
   \frac{\mbox{Br}(\pi^+\pi^-)}{\mbox{Br}(\pi^\mp K^\pm)}
   &=& 0.5\mbox{--}1.9 \quad [0.25\pm 0.10] \,, \nonumber\\
   \frac{\mbox{Br}(\pi^\mp K^\pm)}{2\mbox{Br}(\pi^0 K^0)}
   &=& 0.9\mbox{--}1.4 \quad [0.59\pm 0.27] \,, \nonumber\\
   \frac{2\mbox{Br}(\pi^0 K^\pm)}{\mbox{Br}(\pi^\pm K^0)}
   &=& 0.9\mbox{--}1.3 \quad [1.27\pm 0.47] \,, \nonumber\\
   \frac{\tau_{B^+}}{\tau_{B^0}}\,
   \frac{\mbox{Br}(\pi^\mp K^\pm)}{\mbox{Br}(\pi^\pm K^0)}
   &=& 0.6\mbox{--}1.0 \quad [1.00\pm 0.30] \,.
\end{eqnarray}
The first ratio is clearly in disagreement with current CLEO 
data~\cite{C00} shown in square brackets. However, there is good 
agreement with the recent results~\cite{Osaka} reported by BaBar 
($0.74\pm 0.29$) and Belle ($0.36\pm 0.26$).

The near equality of the second and third ratios in (\ref{rats}) is 
a consequence of isospin symmetry.\cite{NR98} We find 
$\mbox{Br}(B\to\pi^0 K^0)=(4.5\pm 2.5)\times 10^{-6}\times 
(|V_{cb}|/0.04)^2\times[F_+^{B\to\pi}(0)/0.3]^2$ almost independently 
of 
$\gamma$. This is three time smaller than the central value reported 
by CLEO~\cite{C00}, $(14.6_{\,-5.1-3.3}^{\,+5.9+2.4})\times 10^{-6}$, 
and four times smaller than the central value reported by 
BELLE~\cite{Osaka}, $(21.0_{\,-7.8-2.3}^{\,+9.3+2.5})\times 10^{-6}$. 
It will be interesting to follow how the comparison between data and 
theory will develop as the data become more precise.

\begin{figure}[t]
\epsfxsize=11cm
\centerline{\epsffile{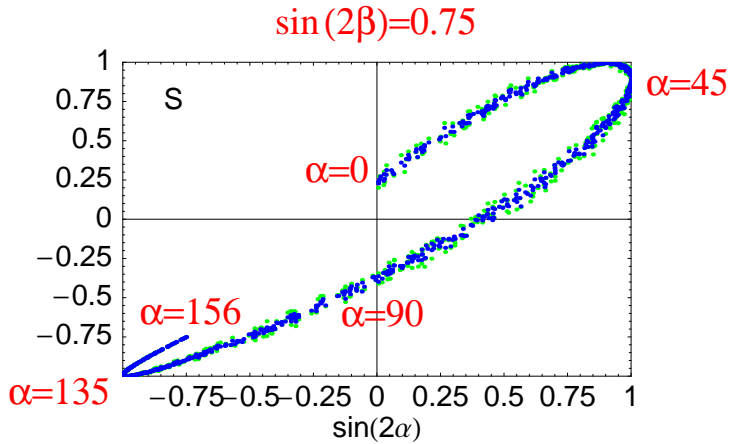}}
\vspace{-0.5cm}
\caption{Mixing-induced CP asymmetry in $B\to \pi^+\pi^-$ decays. 
The lower band refers to values $45^\circ<\alpha<135^\circ$, the
upper one to $\alpha<45^\circ$ (right) or $\alpha>135^\circ$ (left). 
We assume $\alpha,\beta,\gamma\in [0,180^\circ]$.}
\label{fig2}
\end{figure}

\subsection{CP asymmetry in $B\to\pi^+\pi^-$ decay}

The stability of the prediction for the $B\to\pi^+\pi^-$ amplitude 
suggests that the CKM angle $\alpha$ can be extracted from the 
time-dependent mixing-induced CP asymmetry in this decay mode, 
without using isospin analysis. Figure~\ref{fig2} displays the 
coefficient $S$ of $-\sin(\Delta m_{B_d} t)$ as a function of 
$\sin(2\alpha)$ for $\sin(2\beta)=0.75$. For some values of $S$ there
is a two-fold ambiguity (assuming all angles are between $0^\circ$ 
and $180^\circ$). A consistency check of the approach could be 
obtained, in principle, from the coefficient of the 
$\cos(\Delta m_{B_d} t)$ term, which is given by the direct CP 
asymmetry in this decay.

\section{Summary and Outlook}

With the recent commissioning of the $B$ factories and the planned 
emphasis on heavy-flavor physics in future collider experiments, the 
role of $B$ decays in providing fundamental tests of the Standard Model 
and potential signatures of New Physics will continue to grow. In many 
cases the principal source of systematic uncertainty is a theoretical 
one, namely our inability to quantify the nonperturbative QCD effects 
present in these decays. This is true, in particular, for almost all 
measurements of direct CP violation. Our work provides a rigorous 
framework for the evaluation of strong-interaction effects for a large 
class of exclusive, two-body nonleptonic decays of $B$ mesons. It 
gives a well-founded field-theoretic basis for phenomenological studies 
of exclusive hadronic $B$ decays and a formal justification for the 
ideas of factorization.

It is our belief that the factorization formula (\ref{fff}) and its
generalization to decays into two light mesons will form a useful 
basis
for future phenomenological studies of nonleptonic $B$ decays. We 
stress, however, that a considerable amount of conceptual work remains 
to be completed. Theoretical investigations along the lines discussed
here should be pursued with vigor. We are confident that, ultimately, 
this research will result in a {\em theory\/} of nonleptonic $B$ 
decays.

\section*{Acknowledgments}
It is a pleasure to thank M.~Beneke, G.~Buchalla and C.~Sachrajda for 
an ongoing collaboration on the subject of this talk. This work was 
supported in part by the National Science Foundation.


\section*{References}

\begin{thebibliography}{99}

\bibitem{BBNS}
M. Beneke, G. Buchalla, M. Neubert and C.T. Sachrajda, 
{\em Phys.\ Rev.\ Lett.} {\bf 83}, 1914 (1999).

\bibitem{bigpaper}
M. Beneke, G. Buchalla, M. Neubert and C.T. Sachrajda, 
Preprint {\tt hep-ph/0006124}.

\bibitem{Bj89}
J.D. Bjorken, 
{\em Nucl.\ Phys.\ B (Proc.\ Suppl.)} {\bf 11}, 325 (1989).

\bibitem{DG91}
M.J. Dugan and B. Grinstein, 
{\em Phys.\ Lett.\ B} {\bf 255}, 583 (1991).

\bibitem{BBNSnew}
M. Beneke, G. Buchalla, M. Neubert and C.T. Sachrajda, 
Preprint {\tt hep-ph/0007256}.

\bibitem{Martin}
M. Beneke, these Proceedings. 

\bibitem{PW91}
H.D. Politzer and M.B. Wise, 
{\em Phys.\ Lett.\ B} {\bf 257}, 399 (1991).

\bibitem{Rodr97}
J.L. Rodriguez (CLEO Collaboration), 
Proceedings of the 2nd International Conference on {\em B Physics 
and CP Violation}, edited by T.E.~Browder et al.\ (World Scientific,
Singapore, 1998), pp.~124 ({\tt hep-ex/9801028}).

\bibitem{CLEO9701}
B. Barish et al.\ (CLEO Collaboration), 
Preprint CLEO~CONF~97-01 (EPS~97-339).

\bibitem{PDG}
C. Caso et al.\ (Particle Data Group), 
{\em Eur.\ Phys.\ J.\ C} {\bf 3}, 1 (1998).

\bibitem{DZ00}
D. Du, D. Yang and G. Zhu, Preprint {\tt hep-ph/0005006}.

\bibitem{Muta}
T. Muta, A. Sugamoto, M. Yang and Y. Yang, Preprint 
{\tt hep-ph/0006022}.

\bibitem{KLS00}
Y.Y. Keum, H.-n.\ Li and A.I. Sanda, 
Preprints {\tt hep-ph/0004004} and {\tt hep-ph/0004173}.

\bibitem{FM98}
R. Fleischer and T. Mannel, 
{\em Phys.\ Rev.\ D} {\bf 57}, 2752 (1998).

\bibitem{NR98}
M. Neubert and J.L. Rosner, 
{\em Phys.\ Lett.\ B} {\bf 441}, 403 (1998);\\ 
M. Neubert, 
{\em JHEP} {\bf 02}, 014 (1999).

\bibitem{NR}
M. Neubert and J.L. Rosner, 
{\em Phys.\ Rev.\ Lett.} {\bf 81}, 5076 (1998);
M. Neubert, 
{\em Nucl.\ Phys.\ B (Proc.\ Suppl.)} {\bf 86}, 477 (2000).

\bibitem{C00}
D. Cronin-Hennessy et al.\ (CLEO Collaboration),
Preprint\\ {\tt hep-ex/0001010}.

\bibitem{Osaka}
Talks by presented by 
T.J.~Champion (BaBar Collaboration) and P.~Chang (Belle Collaboration) 
at the XXXth International Conference on High-Energy Physics, 
27~July -- 2~August, 2000, Osaka, Japan.

\end{thebibliography}
\end{document}